\pdfoutput=1
\documentclass[aps,prl,twocolumn,superscriptaddress,a4paper,
    floatfix]{revtex4-1}
\usepackage[latin1]{inputenc}
\usepackage{graphicx}
\usepackage{amsmath,amssymb}
\usepackage{hyperref}
\usepackage{color}
\usepackage{ulem} 
\usepackage{floatrow}

\usepackage{braket}
\usepackage{gensymb}

\begin{document}

\preprint{APS/123-QED}

\title{Coupling two order parameters in a quantum gas}
\author{Andrea Morales}
\affiliation{Institute for Quantum Electronics, ETH Zurich, 8093 Zurich, Switzerland}

\author{Philip Zupancic}
\affiliation{Institute for Quantum Electronics, ETH Zurich, 8093 Zurich, Switzerland}

\author{Julian L\'eonard}
\affiliation{Institute for Quantum Electronics, ETH Zurich, 8093 Zurich, Switzerland}
\affiliation{Department of Physics, Harvard University, Cambridge, Massachusetts 02138, USA.}

\author{Tilman Esslinger}
\email{esslinger@phys.ethz.ch}
\affiliation{Institute for Quantum Electronics, ETH Zurich, 8093 Zurich, Switzerland}

\author{Tobias Donner}
\affiliation{Institute for Quantum Electronics, ETH Zurich, 8093 Zurich, Switzerland}

\maketitle
{\bf Controlling matter to simultaneously support multiple coupled properties is of fundamental and technological importance \cite{Spaldin2010}. For example, the simultaneous presence of magnetic and ferroelectric orders in multiferroic materials leads to enhanced functionalities \cite{Cheong2007, Hur2004, Lawes2005, Heyer2006}. In high-temperature superconductors, intertwining between charge- and spin-order can form superconducting states at high transition temperatures \cite{Tranquada1995, Carbotte1999, daou2010, Hinkov2008}. However, pinning down the microscopic mechanisms responsible for the simultaneous presence of different orders is difficult, making it hard to predict the phenomenology of a material \cite{DemlerSO5, Artyukhin2014} or to experimentally modify its properties \cite{Norman98, Hill2000, Aschauer2014, Tsvelik2007, Nyeki2017}. Here we use a quantum gas to engineer an adjustable interaction at the microscopic level between two orders, and demonstrate scenarios of competition, coexistence and coupling between them. In the latter case, intriguingly, the presence of one order lowers the critical point of the other. Our system is realized by a Bose-Einstein condensate which can undergo self-organization phase transitions in two optical resonators \cite{Leonard2016}, resulting in two distinct crystalline density orders. We characterize the intertwining between these orders by measuring the composite order parameter and the elementary excitations. We explain our results with a mean-field free energy model, which is derived from a microscopic Hamiltonian. Our system is ideally suited to explore properties of quantum tricritical points as recently realized in \cite{Friedemann2017} and can be extended to study the interplay of spin and density orders \cite{Mivehvar2017} also as a function of temperature \cite{Piazza2013}.
}

\begin{figure}[h!]
	\centering
		\includegraphics[width=\columnwidth]{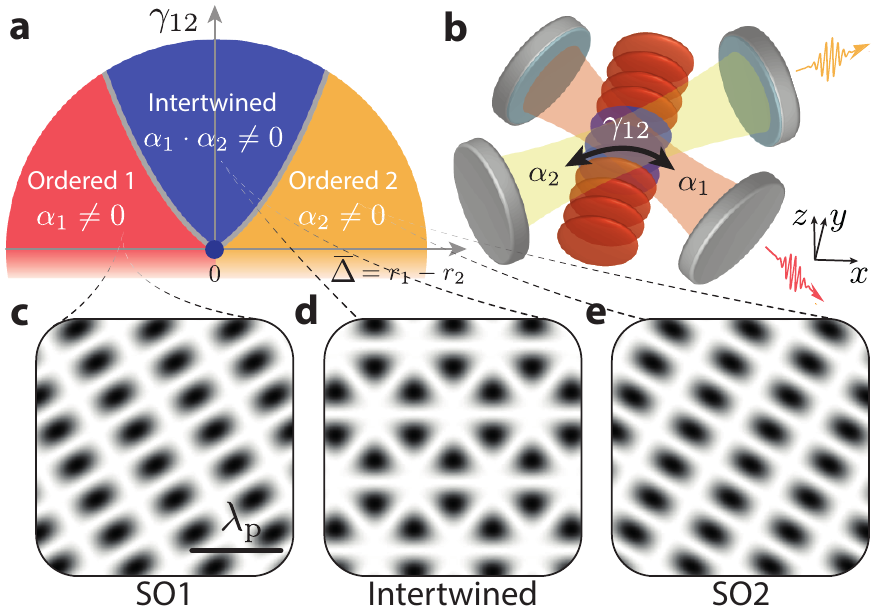}
	\caption{ \textbf{Intertwined order of two order parameters.} {\bf a,} Phase diagram of a system exhibiting two different ordered phases associated to the order parameters $\alpha_1$ and $\alpha_2$ (red and yellow) as a function of the imbalance parameter $\bar{\Delta} = r_1-r_2$. The existence of a direct coupling term $\gamma_{12}>0$ between $\alpha_1$ and $\alpha_2$ leads to the formation of an additional phase with intertwined order (blue region). For $\gamma_{12}=0$ at $\bar{\Delta}=0$ a special point with enhanced symmetry is present (dark blue dot). {\bf b,} A Bose-Einstein condensate (blue) is placed at the intersection of two optical cavities crossing under an angle of $60^\circ$. It is illuminated along $y$ by a standing wave laser beam (dark red) which is tilted by $60^\circ$ with respect to both cavities and is referred to as transverse pump. Crystallization of the BEC can occur via scattering of photons from the transverse pump into one of the two cavities or into both. As a consequence, the cavity light field amplitudes $\alpha_1$ and $\alpha_2$ can acquire non zero values. The photons leaking out of the cavities are detected with single photon counters. The corresponding orders intertwine with each other via the cavity-cavity scattering rate $\gamma_{12}$. {\bf c, d, e}  Calculated density modulations associated to density orders mediated by photon scattering into cavity 1 (SO1), cavity 2 (SO2) and both cavities simultaneously (intertwined). White (black) regions are associated to low (high) atomic densities. The wavelength of a transverse pump photon is $\lambda_\textrm{p}\approx 780\,\;$nm.}
	\label{fig:scheme}
\end{figure}
In Landau theory, the phenomenology of continuous phase transitions is described by an order parameter that minimizes the free energy $\mathcal{F}$ and that is zero in the normal phase and finite in the ordered phase \cite{Sachdev1999}. In the case of a system with two order parameters $\alpha_1$, $\alpha_2$ the minimal description capturing the possible phases is given by the free energy $\mathcal{F}(\alpha_1,\alpha_2)/\hbar = r_1\alpha_1^2+r_2\alpha_2^2+g(\alpha_1^2+\alpha_2^2)^2-\gamma_{12} \alpha_1^2\alpha_2^2$, where $g>0$, $r_1$, $r_2$ and $\gamma_{12}$ are the couplings that are determined by the microscopic details of the interactions. The resulting phases depend on the values of $\gamma_{12}$, $r_1$ and $r_2$ (see Fig.~\ref{fig:scheme}{\bf a}). If $\gamma_{12}<0$, minimizing $\mathcal{F}$ always requires $\alpha_1=0$ or $\alpha_2=0$, i.e. the order parameters mutually exclude each other. If $\gamma_{12}>0$, however, $\mathcal{F}$ supports minima with $\alpha_1\cdot\alpha_2\neq0$ for certain ranges of $r_1$, $r_2$ and $g$, and the orders can intertwine \cite{Fradkin2015}. The special case of negligible $\gamma_{12}$ supports a phase transition with enhanced symmetry \cite{Leonard2016}.

We report on the realization of a system where we control the microscopic process that favours intertwined order between two different density patterns. Our experimental realization is illustrated in Fig.~\ref{fig:scheme}{\bf b}. A Bose-Einstein condensate (BEC) of N=2.5(1)$\times10^5$ $^{87}$Rb atoms is optically trapped at the intersection of two optical cavity modes. The BEC is illuminated by a standing wave laser beam with frequency $\omega_\textrm{p}$, which we refer to as transverse pump. It is far detuned, by $\Delta_\textrm{A}=\omega_\textrm{p}-\omega_\textrm{A}<0$, from the D$_2$ atomic transition $\omega_\textrm{A}$ and closely detuned, by $\Delta_i=\omega_\textrm{p}-\tilde\omega_i<0$, from the dispersively shifted cavity resonance frequencies $\tilde\omega_i=\omega_i+Ng_i^2/(2\Delta_\textrm{A})$, $i\in \{1,2\}$ (see Methods). 

\begin{figure*}[t]
	\floatbox[{\capbeside\thisfloatsetup{capbesideposition={right,top},capbesidewidth=50mm}}]{figure}[\FBwidth]
	{\includegraphics[width=115mm]{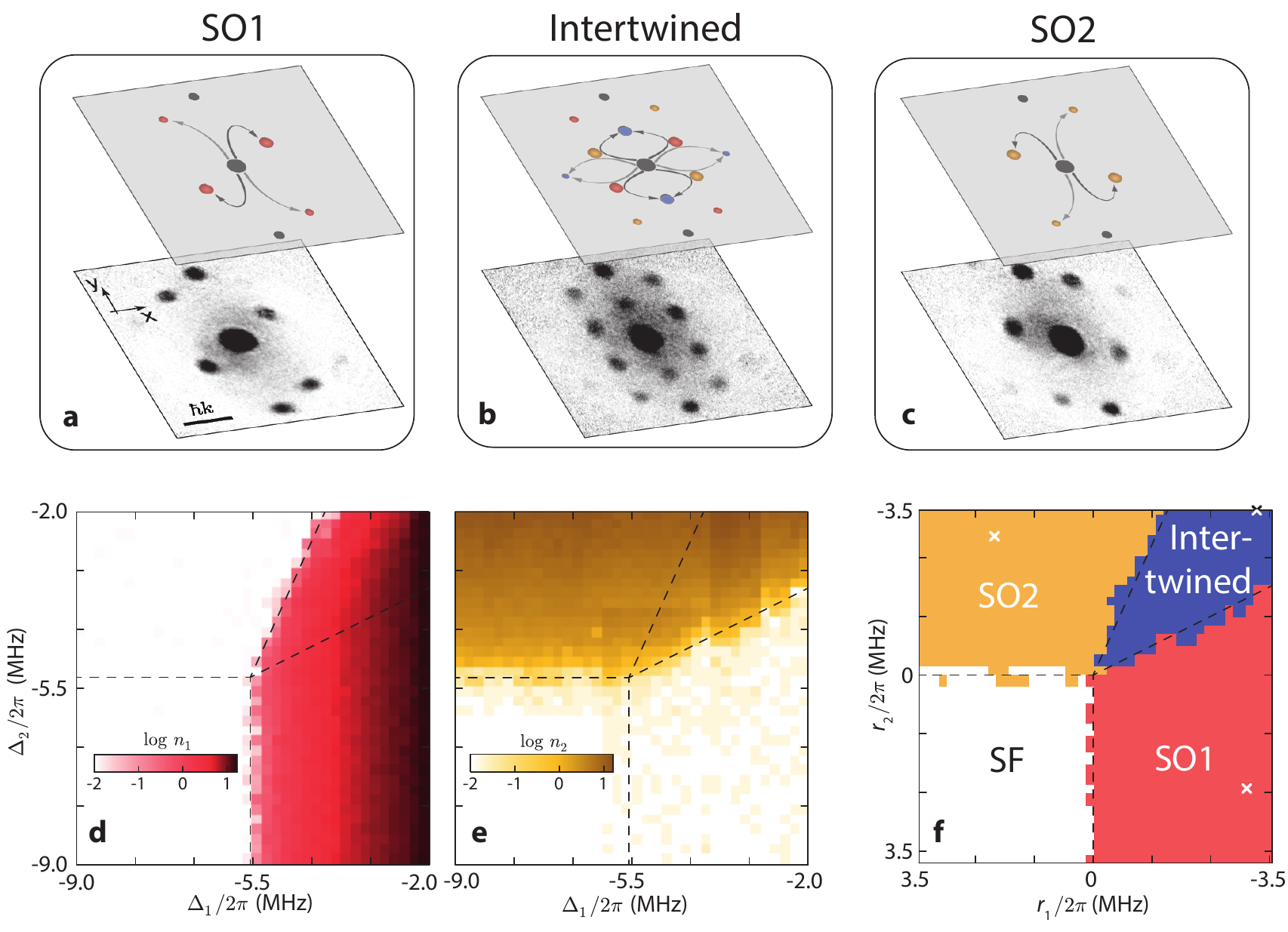}}
	{\caption{ \textbf{Observing intertwined order.} {\bf a,b,c,} Imaging the atoms after ballistic expansion reveals their momentum distributions (lower panels), reflecting the scattering processes (see Methods) between cavities and transverse pump (upper panel). These distributions are the Fourier transforms of the modulated atomic density shown in Fig.~1. Dark regions indicate high atomic densities. {\bf d,e,} Simultaneously measured intracavity photon numbers $n_1$ (red) and $n_2$ (yellow), as a function of the respective detunings $\Delta_1$ and $\Delta_2$ from the transverse pump beam. Dashed lines are the fitted phase boundaries. {\bf f,} We identify four different regions (see Methods): a superfluid without photons (SF), two regions with photons in only one of the cavities (SO1/SO2), and a region where photons populate both cavities simultaneously (Intertwined). Crosses mark the settings of the measurements in \textbf{a-c} and manifest the relation between light fields and density order.}	\label{intertwined:phase}}
\end{figure*}

An effective atom-atom interaction originates from Raman processes between transverse pump and cavity. In an intuitive picture, a transverse pump photon is off--resonantly scattered by one atom of the BEC into either of the two cavity modes. The cavity photon can then be scattered back into the pump beam by any other atom. During these processes atoms are recoiling from the BEC state into non-zero momentum states with opposite momenta, which are sketched in the upper panels of Fig.~\ref{intertwined:phase}{\bf a}, {\bf c}. In real-space their interference gives rise to a density ripple on the BEC. While the occupation of these momentum modes by the atoms increases the kinetic energy of the system, a photon scattered into either of the cavities creates a lattice potential that lowers the potential energy. A self-organization phase transition to a spatially ordered state occurs beyond a critical point, when the potential energy associated with the formation of a coherent cavity field with amplitude $\alpha_i=\pm\sqrt{n_i}$, overcomes the kinetic energy of the atoms \cite{Baumann2010}. The variable $n_i$ denotes the mean intracavity photon number. As a result, the superfluid acquires a chequerboard density structure (see Fig.~\ref{fig:scheme}{\bf c} and {\bf e}). Depending on the sign of $\alpha_i$, the density maxima can be located either on the even or on the odd sites of the underlying chequerboard potential, corresponding to a broken $\mathbb{Z}_2$ symmetry \cite{Baumann2011}.

Intertwining between the two order parameters is mediated by an additional microscopic mechanism, the scattering of photons from one cavity to the other via the atoms, and favours the formation of another density structure on the superfluid  (see Fig.~\ref{fig:scheme}{\bf d}). These scattering processes couple the BEC to additional momentum states (see upper panel in Fig.~\ref{intertwined:phase}{\bf b}).

The Hamiltonian description of our system maps onto the free energy $\mathcal{F}(\alpha_1,\alpha_2)$, with the parameters (see Methods)
\begin{align}
	\begin{split}
		r_i=&-\Delta_i\Bigl(1-\frac{\lambda^2}{\lambda_{i,\textrm{crit}}^2}\Bigl), \,\;  i\in \{1,2\}\\
		g=&\frac{8\lambda^4}{\omega_\textrm{r}^3}, 	\,\;  \,\;	\gamma_{12}=\frac{4\lambda_{12}^2}{\omega_\textrm{r}}.\\
	\end{split}
\label{eq:free energy}
\end{align}

Here $\omega_\textrm{r}$ = $2\pi \times$3.7\,kHz is the recoil frequency of the atoms for a transverse pump photon, the parameter $\lambda=-\sqrt{N}\Omega_\textrm{p}g_0/(2\sqrt{2}\Delta_\textrm{A})$ is the Raman coupling associated to scattering processes between the pump field -- with Rabi frequency $\Omega_\textrm{p}$ -- and either cavity field -- with vacuum-Rabi frequencies $g_0$ (see Methods), $\lambda_{i,\textrm{crit}}=\sqrt{-\Delta_i\omega_\textrm{r}/4}$ is the critical coupling at which self-organization in cavity $i$ occurs and $\lambda_{12}=-Ng_0^2/(2\sqrt{2}\Delta_\textrm{A})$ is the Raman coupling associated to scattering of photons from one cavity to the other.

In our experiment we can independently control the four coefficients in  \eqref{eq:free energy}. The Raman coupling $\lambda$ can be controlled by the power $P$ of the transverse pump as $\Omega_\textrm{p}\propto\sqrt{P}$, while the critical couplings $\lambda_{i,\textrm{crit}}$ can be changed via the detunings $\Delta_i$. The coefficient $\lambda_{12}$ is controlled by the atomic detuning $\Delta_\textrm{A}$.

The first goal of our experiment is to determine the phases of the system. Our experimental observable is the Fourier transform of the BEC density structure, which we measure by switching off the optical trap and performing absorption imaging of the cloud after ballistic expansion \cite{Ketterle1999}. In this initial experiment, the coefficient $\gamma_{12}$ is fixed to a non-zero value by choosing $\Delta_\textrm{A}/2\pi$ = -1\,THz. We explore three different scenarios. In the first two, we set a large positive or negative detuning imbalance $|\bar\Delta|=|\Delta_2-\Delta_1|$, by setting either $\tilde\omega_\textrm{1}$ significantly closer to $\omega_\textrm{p}$ than  $\tilde\omega_\textrm{2}$, or vice versa. In terms of the free energy $\mathcal{F}$, this situation corresponds to having $r_1$ and $r_2$ significantly different from each other and therefore either $\alpha_1$ or $\alpha_2$ dominates. In the third scenario, we set $\bar\Delta\sim0$, therefore maintaining $r_1\sim r_2$. After having fixed the cavity detunings $\Delta_\textrm{i}$, we induce crystallization of the BEC by increasing the transverse pump power within 100 ms to the same final value of $10(1)\hbar\omega_\textrm{r}$ in each run. This ensures a constant $\lambda$ in all the three measurements. Three qualitatively different momentum distributions of the BEC are visible, corresponding to self-organization of the BEC in cavity 1 (SO1), cavity 1 and 2 (Intertwined) or cavity 2 (SO2) (see lower panels Fig.~\ref{intertwined:phase}{\bf a-c}). These signal the existence of different crystalline orders in the system. In particular, the additional momentum peaks occupied in the intertwined phase (blue peaks in the upper panel of Fig.~\ref{intertwined:phase}{\bf b}) are representative of the fact that the `parent' orders do not just add to each other but they give rise to a more complex spatial arrangement. For the three measured Fourier transforms of the crystal structure  (Fig.~\ref{intertwined:phase}{\bf a-c}) we observed photon occupation either in cavity 1, in both cavities, or in cavity 2, by recording the photons leaking from the cavities onto single photon detectors .

We map out the phase diagram using the connection between the presence of the cavity light fields and the existence of a certain crystalline order in the BEC. We repeat the previous experiments at different pairs ($\Delta_1$, $\Delta_2$), covering positive and negative values for $r_1$ and $r_2$. For each pair ($\Delta_1$, $\Delta_2$) we measure the steady state intracavity photon numbers (see Fig.~\ref{intertwined:phase}{\bf d}, {\bf e}). 

The resulting phase diagram is reported in Fig.~\ref{intertwined:phase}{\bf f}. For positive $r_1$ and $r_2$, corresponding to the white area in the diagram, there is no light in the cavities and the minima of $\mathcal{F}$ are located at $\alpha_1=\alpha_2=0$. One phase boundary is located on the line $r_1=0$, defined by $\lambda=\lambda_{1,\textrm{crit}}$. Beyond this line (red/SO1 region in Fig.~\ref{intertwined:phase}{\bf f}), $r_1$ is negative and the order parameter $\alpha_1$ becomes non--zero in the ground state of the system. Similarly for the yellow region (SO2), beyond the line $r_2=0$ of the second phase boundary, $r_2$ is negative and $\alpha_2$ is non--zero. The two phase boundaries cross each other when $r_1=r_2=0$ and a multicritical point is present. We also observe that the two order parameters compete, i.e. the critical points are shifted to higher values. This can be seen in the region where both $r_1$ and $r_2$ are negative, where a change in the slope of the phase boundaries is visible for both phases. We can identify regions of mutually exclusive orders, where only one of the two $\alpha_i$ is non--zero even though both $r_1, r_2<0$. Additionally, we observe a region where both order parameters are non--zero (blue region in Fig.~\ref{intertwined:phase}{\bf f}), a signature of a phase with intertwined order.

\begin{figure}
	\centering
		\includegraphics[width=\columnwidth]{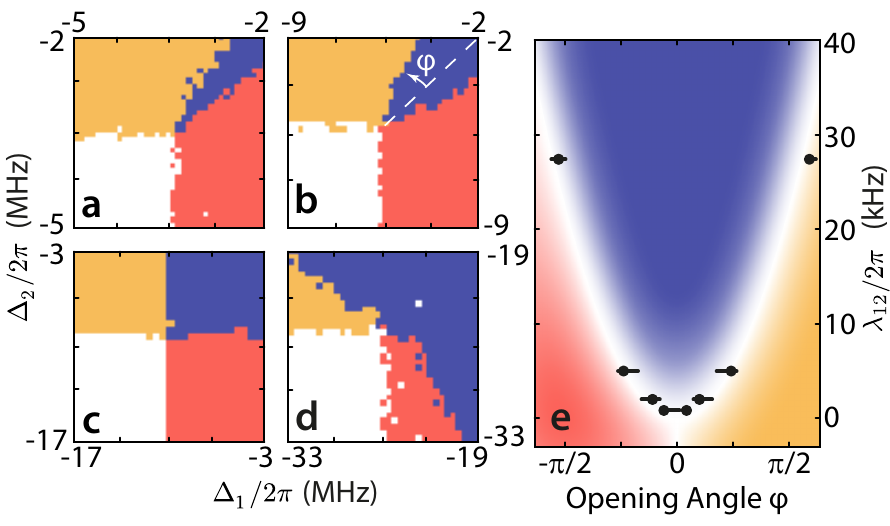}
	\caption{ \textbf{Controlling intertwined order.}  {\bf a-d,} Phase diagrams extracted from the order parameters $\alpha_1$ and $\alpha_2$ measured as in Fig.~\ref{intertwined:phase}{\bf d} but at different $\Delta_\textrm{A}/2\pi=-2.4$\,THz ({\bf a}), $-1$\,THz ({\bf b}),$-0.4$\,THz ({\bf c}) and $-0.073$\,THz ({\bf d}).  Four phases are visible, associated with no light in either cavity (white), light in cavity 1 (red), light in cavity 2 (orange) and light in both cavities (blue). The latter region gets larger as we approach the atomic resonance. {\bf e,} The opening angle $\varphi$ of the intertwined phase as defined in \textbf b is used to construct the phase diagram of the intertwined order parameters as a function of $\lambda_{12}$. The errorbars reflect the uncertainty in fitting the phase boundary, taking into account a freedom of choice of the photon number threshold (see Methods).}	
	\label{intertwined:phase diagram}
\end{figure}

The extent of the phase with intertwined order depends on the coefficient $\gamma_{12}$. We can control it by changing the atomic detuning $\Delta_\textrm{A}$ either to smaller absolute values (larger $\gamma_{12}$) or larger absolute values (smaller $\gamma_{12}$). By tuning $\gamma_{12}$ towards zero, we can favour a regime of mutual exclusion, i.e. competition of the two orders (see Fig.~\ref{intertwined:phase diagram}{\bf a}). Here, the phase with intertwined order shrinks to a narrow region where a $U(1)$--symmetry in the free energy emerges \cite{Leonard2016, Leonard2017, Zwerger2017}. Alternatively, by fine tuning to $\gamma_{12}=2g$ we realize a situation where the free energy $\mathcal{F}$ separates (see Methods) into the sum of two independent free energies, each associated to one of the two order parameters. In this case the order parameters coexist, i.e. they do not influence each other throughout the entire phase diagram (see Fig.~\ref{intertwined:phase diagram}{\bf c}). Finally, we can also access the regime where $\gamma_{12}>2g$ (see Fig.~\ref{intertwined:phase diagram}{\bf d}), i.e. the order parameters are coupled. In this case, intringuingly, their critical points are shifted towards lower values by the coefficient $\gamma_{12}$. We use the opening angle $\varphi$ shown in Fig.~\ref{intertwined:phase diagram}{\bf b} to quantify the extent of the phase with intertwined order and construct the phase diagram as a function of $\lambda_{12}$ (see Fig.~\ref{intertwined:phase diagram}{\bf e}).

\begin{figure}[t]
	\centering
		\includegraphics[width=\columnwidth]{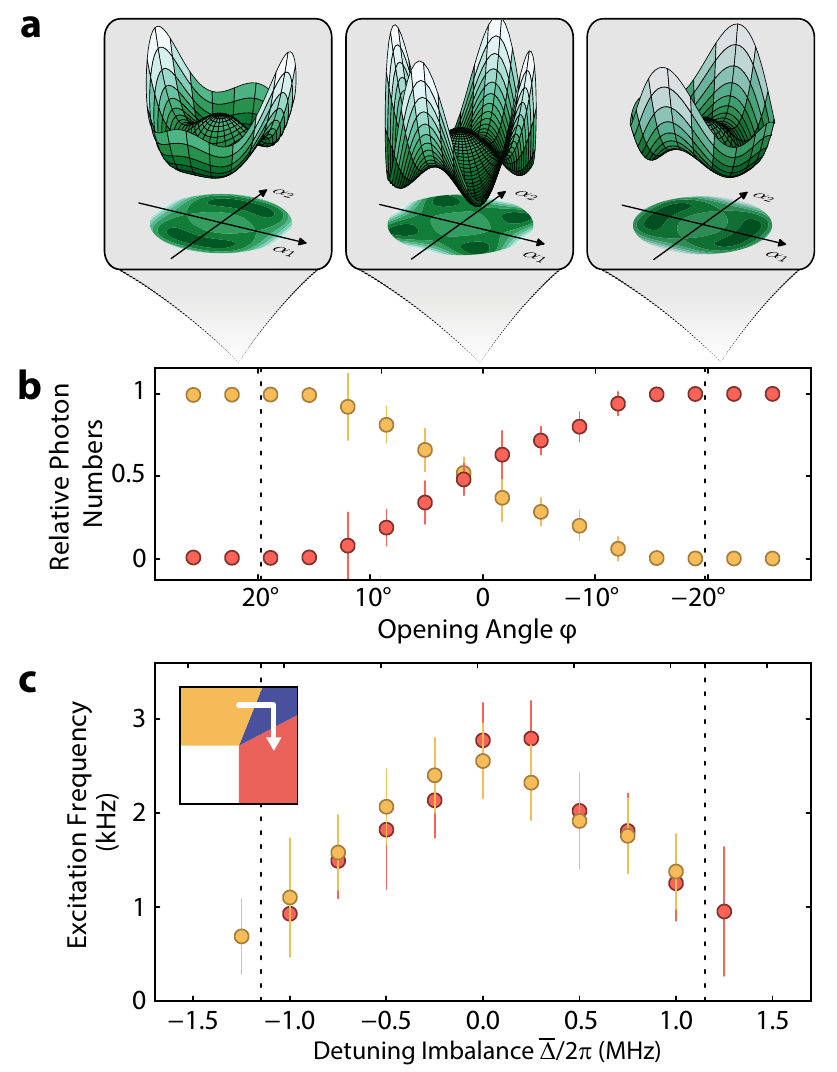}
	\caption{ \textbf{Characterization of intertwined order.} {\bf a,} Landau free energy in the phase SO1 (right), the intertwined phase (center), and the phase SO2 (left). The two axes of the horizontal plane are defined by the cavity light field amplitudes $\alpha_1$ and $\alpha_2$. When crossing the phase boundary (dashed lines) to the intertwined phase, the two ground states split into two, each from a single minimum located along one of the two axes. The positions of the two minima rotate on a circle in opposite directions when the detuning imbalance $\bar\Delta$ is changed. {\bf b,} The rotation is signaled experimentally by a smooth change in the photon numbers in the two cavities. The data is extracted from Fig.~2 \textbf{d,e} (see Methods). {\bf c,} The azimutal curvature of the Landau potential also changes. We measure this curvature by probing the excitation frequency of the azimuthal mode in the intertwined phase. Vertical error bars reflect standard deviations and fit uncertainties (see Methods). $\bar\Delta$ has a preparation uncertainty of 50\,kHz. All the data are taken at $\Delta_\textrm{A}/2\pi=-1$\,THz.}
	\label{intertwined:admixture}
\end{figure}

Having shown that intertwined order can go beyond the mere coexistence of the order parameters, we characterize the deformation of the free energy as a function of the detuning imbalance $\bar\Delta$, when $r_1, r_2<0$ (see Fig.~\ref{intertwined:admixture}). As the system enters the intertwined phase, each of its two ground state minima splits into two. Therefore, a $\mathbb{Z}_2\times\mathbb{Z}_2$ symmetry is broken (see central panel). As a function of the detuning imbalance $\bar\Delta$, the positions of the four minima change. Experimentally, this change can be revealed by extracting the photon numbers $\bar{n}_1$, $\bar{n}_2$ as a function of the angle $\varphi$ from Fig.~\ref{intertwined:phase}{\bf d},{\bf e}. This is shown in Fig.~\ref{intertwined:admixture}{\bf b}. When $\varphi$ is tuned from positive to negative values, the photon numbers exchange smoothly. 

To further confirm our model of tunable symmetry, we probe the curvature of the free energy by equivalently measuring the excitation frequency for different $\bar\Delta$ at $\Delta_\textrm{A}/2\pi=-1$\,THz (see Fig.~\ref{intertwined:admixture}{\bf c}). We apply a variant of Bragg spectroscopy that we introduced in a previous work \cite{Leonard2017}. Here additional probe light of variable frequency different from the transverse pump, is injected on axis in either cavity. The excitation energy is then identified as the detuning corresponding to a sudden change in the intracavity photon number (see Methods). Starting at $\bar\Delta=0$, we observe that the excitation energy decreases approaching the phase boundaries (see Fig.~\ref{intertwined:admixture}{\bf c}). These measurements are in agreement with the expected mode softening at a second order phase transition from a phase with broken $\mathbb{Z}_2$-- to a phase with broken $\mathbb{Z}_2\times\mathbb{Z}_2$--symmetry (see Methods). 

In future, this experimental platform can be modified to simulate coupled spin and charge order reminiscent of high-temperature superconductors \cite{Mivehvar2017}, or to realize the Coleman-Weinberg model by setting $\gamma_{12}<0$ \cite{Bornholdt1995}, or to realize phases with vestigial order \cite{Gopalakrishnan2017}. In addition, the photons leaking from the cavities allow to access the fluctuations of the order parameter and their scaling when approaching a critical point to confirm theoretical models as presented in the accompanying paper \cite{Awadhesh2017}.


\section*{METHODS}

\noindent \textbf{Setup and preparation of the Bose-Einstein condensate (BEC)} We prepare an almost pure BEC of $N=2.5(1)\times 10^5$ atoms in an optical dipole trap formed by two orthogonal laser beams at a wavelength of $1064\,\text{nm}$ along the $x$-- and $y$--axes. The trapping frequencies are $(\omega_x, \omega_y, \omega_z)=2\pi \times (66(1), 75(1), 133(5))\,\mathrm{Hz}$. The trap position coincides with the crossing point of the fundamental Gaussian modes of the two optical cavities with Rabi frequencies $(g_1, g_2) = 2\pi\times (1.95(1), 1.77(1))\,\mathrm{MHz}$, and decay rates $(\kappa_1, \kappa_2) = 2\pi\times (147(4), 800(11))\,\mathrm{kHz}$. We can adjust the resonance frequencies of the cavity modes with piezoelectric elements that are included in the mount of each cavity mirror. The frequencies are actively stabilized with the help of an additional laser beam at $830\,\text{nm}$.

\noindent \textbf{Lattice and photon number calibrations} In order to calibrate the lattice depths of the transverse pump and each cavity field we perform Raman-Nath diffraction on
the atomic cloud. The intracavity photon number calibration can then be calculated from the lattice depth per photon $U_0=\hbar g^2_i/\Delta_\mathrm{A}$, where $\hbar$ is the reduced Planck constant, $g_i$ is the vacuum Rabi frequency of the cavity, and $\Delta_\textrm{A}$ is the detuning from atomic resonance. We extract efficiencies of $(5.0(2)\%; 1.4(1)\%)$ for detecting an intracavity photon from cavity $i$ with single-photon counting modules.

\noindent \textbf{Spin-assisted self-organization}  The BEC is prepared in the $|F,m_\textrm{F}\rangle=|1,-1\rangle$ state. All data for the phase diagrams were taken with an offset field of $B_z=55$G, creating a Zeeman level splitting large compared to $\Delta_i$. In this way we suppress collective Raman transitions between different Zeeman sublevels that scatter light in the cavities. 

\noindent \textbf{Evaluation of the phase diagrams}  From the photon data recorded at the cavity output (see Fig.~\ref{intertwined:phase}{\bf d} and {\bf e}), we extract the phase diagram by binarizing the photon data with a photon threshold set to 1 photon/ms. The phase diagram in Fig.~\ref{intertwined:phase}{\bf f} and the ones in Fig.~\ref{intertwined:phase diagram} are calculated in this manner.

\begin{figure}[t]
	\centering
		\includegraphics[width=\columnwidth]{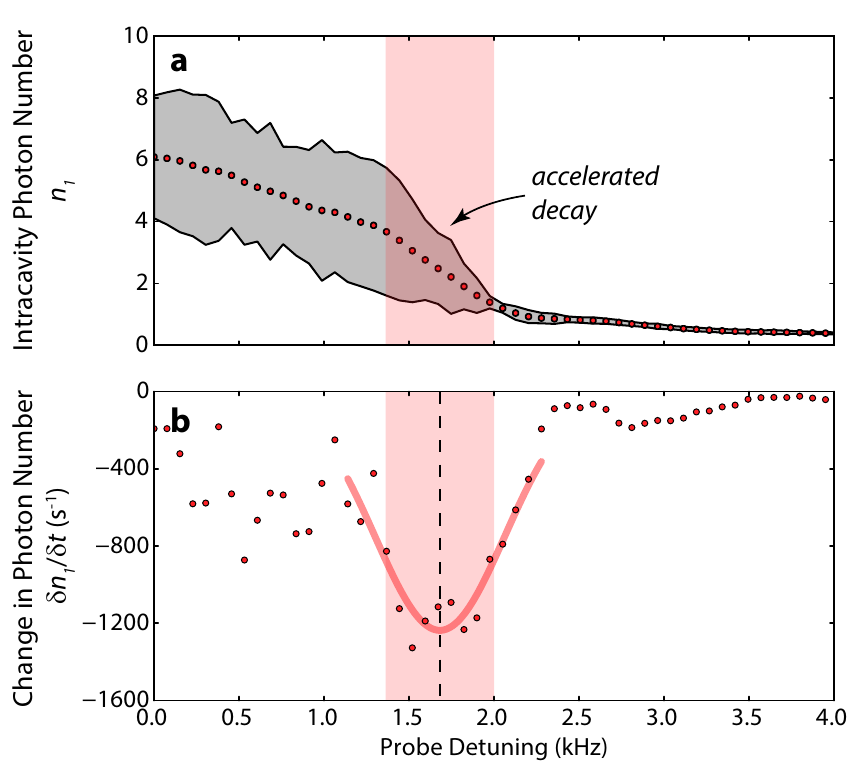}
	\caption{\textbf{Measurement of the excitation energy.} The figure illustrates the measurement for the case of $\Delta_1/2\pi=-4\,\mathrm{MHz}$, $\Delta_2/2\pi=-4.75\,\mathrm{MHz}$, i.e.~$\bar\Delta/2\pi=0.75\,\mathrm{MHz}$, measured on cavity 1. \textbf{a,} Order parameter $n_1$ responding to the weak probe field whose detuning is changing from  0\,kHz to 4\,kHz with respect to the transverse pump frequency within 20\,ms. The data is binned in 0.25ms intervals, smoothened, and averaged over 20 realizations. The gray shaded area shows the standard deviation of the full sample. The red shaded area highlights the region of accelerated decay. \textbf{b,} The time derivative of the mean photon trace shows a resonance at a detuning of 1.7\,kHz (dashed line). A collective mode is excited at this frequency, which subsequently decays. The excess energy of the decay leads to heating which we observe through the decreasing order parameter. We extract the resonance frequency with a local Gaussian fit (light red).}
	\label{fig:excitations}
\end{figure}

\noindent \textbf{Excitation spectrum} To measure the excitation spectrum we exploit a method which we have introduced in a previous work \cite{Leonard2017}. We first fix the cavity detunings from the pump laser frequency to certain values $(\Delta_1^{\mathrm{eq}}, \Delta_2^{\mathrm{eq}})$. We then prepare the BEC at a certain coupling strength by linearly increasing the transverse pump intensity within $50\,\mathrm{ms}$ to a lattice depth of $12.5(10)\,\hbar\omega_\mathrm{rec}$, with $\omega_\mathrm{rec}$ being the recoil frequency for a transverse pump photon. This brings the system to a certain position in the phase diagram where we want to probe the excitations. Subsequently, a small probe field is injected on axis with one of the cavity modes and its detuning is ramped linearly in time from 0\,kHz to 6\,kHz with respect to the transverse pump frequency within 20\,ms. Simultaneously, we measure the response of the system to the probe field by continuously monitoring the number of photons leaking from the cavity mirrors on single photon detectors. We bin photons in 0.25\,ms intervals, apply a smoothing filter and average 20 traces for each position in the phase diagram. When the probe field detuning is on resonance with a polaritonic excitation, the system is driven and the excess energy leads to heating and an accelerated decay of the order parameter, i.e.\ the photon number $n_i$ in cavity $i$. The decay rate $\delta n_i/\delta t$ shows a resonant feature. We extract the excitation energy with a local Gaussian fit. Fig.~\ref{fig:excitations} shows an examplary measurement.

\begin{figure}[t]
	\centering
		\includegraphics[width=\columnwidth]{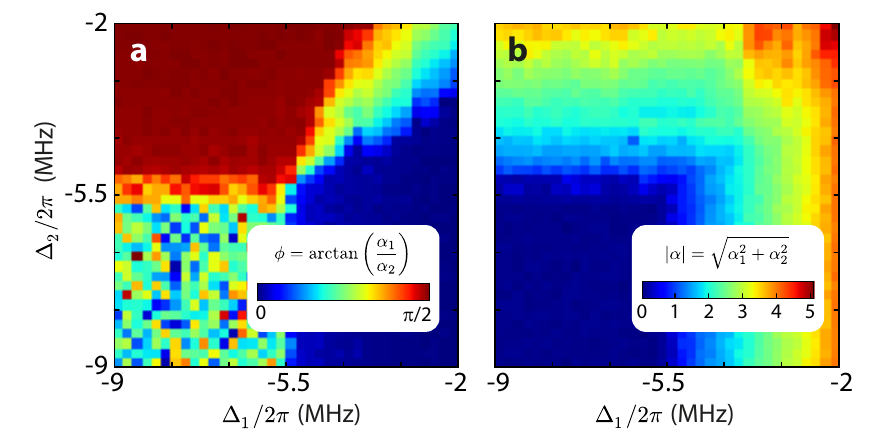}
	\caption{\textbf{Phase and amplitude of the order parameter.} The data from Fig.~2 of the main text is transformed into the phase $\phi=\arctan(\alpha_1/\alpha_2)$ (\textbf{a}) and amplitude $|\alpha|=\sqrt{\alpha_1^2+\alpha_2^2}$ (\textbf{b}) of the composite order parameter. The phase is random in the non-ordered phase ($\Delta_1/2\pi, \Delta_2/2\pi<-5.5\,\mathrm{MHz}$), fixed at 0 or $\pi/2$ in the singly self-organized phase of cavity 1 or 2 respectively, and smoothly varying between them in the intertwined phase. }
	\label{fig:phaseamplitude}
\end{figure}

\noindent \textbf{Intertwined order: amplitude and phase of the order parameter}
From the photon data reported in Fig.~2  of the main text we can extract phase and amplitude of the order parameter making use of the relations $\phi=\arctan(\alpha_1/\alpha_2)$ and $|\alpha|=\sqrt{\alpha_1^2+\alpha_2^2}$. Both are shown in Fig.~\ref{fig:phaseamplitude}. Across the interwined phase, the phase $\phi$ is smoothly changing by $\pi/2$ corresponding to a rotation of the composite order parameter $\alpha_1+i\alpha_2$. From  Fig.~\ref{intertwined:phase}{\bf d,e} we calculate the admixture of the order parameters $\alpha_1$ and $\alpha_2$ across the intertwined phase. This is shown in Fig.~\ref{intertwined:admixture}{\bf b} by averaging photon numbers along rays originating from the multicritical point corresponding to fixed values of the opening angle $\varphi$ that we have introduced in the main text. The photon numbers $n_i$ are normalized to the total photon number and averaged along each ray. Vertical error bars are the standard deviations along the ray. 

\begin{figure}[t]
	\centering
		\includegraphics[width=\columnwidth]{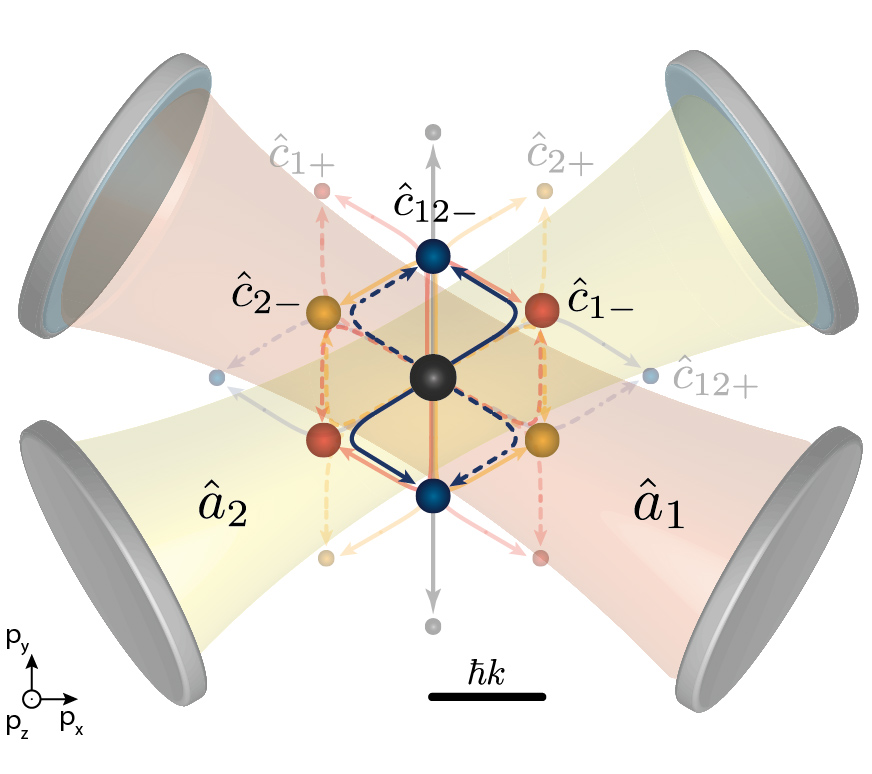}
	\caption{\textbf{Field operators describing the scattering processes.} The cavity modes, with photon annihilation operators $\hat{a}_1$ and $\hat{a}_2$, are sketched in red and yellow. Red (yellow) circles are atomic momentum states populated via photon scattering from the pump into cavity 1 (cavity 2) and vice versa. The associated annihilation operators are $\hat{c}_{1\pm}$ and $\hat{c}_{2\pm}$ respectively. Blue circles are atomic momentum states occupied via photon scattering from one cavity to the other, associated with the annihilation operator $\hat{c}_{12\pm}$. Black circles represent the BEC state and the Bragg peaks of the transverse pump lattice oriented along $y$. The momentum peaks in darker colour have energy $\hbar\omega_-$ whereas those in transparency have higher energy $\hbar\omega_+$ and are neglected in the derivation of the free energy. The scale $\hbar k$ indicates the atomic recoil momentum given by a pump photon. Solid and dahed lines correspond to different time ordered two-photon scattering processes.}
	\label{fig:momenta}
\end{figure}

\noindent \textbf{Hamiltonian description of the system} Raman scattering between the cavity fields (with wave-vector ${\bf k}_i$, $i$ $\in\{1,2\}$) and the transverse pump (with wave-vector ${\bf k}_\textrm{p}$) via the atoms couples the atomic momentum state of the BEC to a superposition state of higher momenta. These states fall into two groups with energies either $\hbar\omega_-=\hbar\omega_{\mathrm{rec}}$ or $\hbar\omega_+=3\hbar\omega_{\mathrm{rec}}$ (see Fig.~\ref{fig:momenta}). The single-photon recoil frequency is given by $\omega_{\textrm{rec}}=\hbar k^2/2m$, with $m$ being the atomic mass and $\hbar$ the reduced Planck constant and $k=|{\bf k}_\textrm{p}|=|{\bf k}_i|$. Following the same procedure described in the Methods section of \cite{Leonard2016}, we include the cavity--cavity interference term and extend the ansatz for the atomic field operator to
\begin{equation}
\begin{aligned}
\hat{\Psi}= &\Psi_0\hat{c}_0+\sum_{i\in \{1,2\}}\bigl(\Psi_{i-}\hat{c}_{i-}+\Psi_{i+}\hat{c}_{i+}\bigl)\\
 &+\Psi_{12-}\hat{c}_{12-} + \Psi_{12+}\hat{c}_{12+},
\end{aligned}
\end{equation}
where $\hat{c}_{i\pm}^\dagger$ ($\hat{c}_{i\pm}$) and  $\hat{c}_0^\dagger$ ($\hat{c}_0$) create (annihilate) an atomic momentum excitation at energy $\hbar\omega_\pm$ associated with cavity $i$ and in the atomic ground state, respectively. $\hat{c}^{\dagger}_{12\pm}$ ($\hat{c}_{12\pm}$) create (annihilate) an atomic momentum excitation at energy $\hbar\omega_{\pm}$ associated to cavity--cavity scattering. $\Psi_0=\sqrt{\frac{2}{A}}$ represents the BEC zero-momentum mode, the functions $\Psi_{i\pm}=\sqrt{\frac{2}{A}}\textrm{cos}[({\bf k}_\textrm{p}\pm{\bf k}_i)\cdot{\bf r}]$ are the atomic modes with momentum imprinted by one of the scattering processes at high-- or low--energy into one of the cavities, and the functions  $\Psi_{12\pm}=\sqrt{\frac{2}{A}}\textrm{cos}[({\bf k}_1\mp{\bf k}_2)\cdot{\bf r}]$ are the atomic modes with momentum imprinted by one of the scattering processes at high-- or low--energy from one cavity to the other. ${\bf r}$ is the real space coordinate vector and $A$ is the area of the Wigner-Seitz cell. We obtain the following effective Hamiltonian
\begin{equation}
\begin{aligned}
\hat{\mathcal{H}} =&\sum_{i\in \{1,2\}}\Bigl[-\hbar\Delta_i\hat{a}^{\dagger}_i\hat{a}_i+\hbar\omega_{+}\hat{c}^{\dagger}_{i+}\hat{c}_{i+}+\hbar\omega_{-}\hat{c}^{\dagger}_{i-}\hat{c}_{i-}\\
&+\frac{\hbar\lambda_i}{\sqrt{N}}\Bigl(\hat{a}^{\dagger}_i+\hat{a}_i\Bigl)\Bigl(\hat{c}^{\dagger}_{i+}\hat{c}_{0}+\hat{c}^{\dagger}_{i-}\hat{c}_{0}+h.c.\Bigl)\Bigl]\\
&+\frac{\hbar\lambda_{12}}{\sqrt{N}}\Bigl(\hat{a}^{\dagger}_1\hat{a}_2+\hat{a}^{\dagger}_2\hat{a}_1\Bigl)\Bigl(\hat{c}^{\dagger}_{12-}\hat{c}_{0}+\hat{c}^{\dagger}_{12+}\hat{c}_{0}+h.c.\Bigl)\\
&+\hbar\omega_{-}\hat{c}^{\dagger}_{12-}\hat{c}_{12-}+\hbar\omega_{+}\hat{c}^{\dagger}_{12+}\hat{c}_{12+},
\end{aligned}
\label{eq:Hamiltonian}
\end{equation}
where the index $i$ labels the two cavities and $N$ is the atom number. $\hat{a}_i^\dagger$ ($\hat{a}_i$) are the creation (annihilation) operators for a photon in cavity $i$.  $\lambda_i=\frac{\eta_i\sqrt{N}}{2\sqrt{2}}$ is the Raman coupling between the transverse pump and cavity $i$, which can be controlled via $\eta_i=-\frac{\Omega_\mathrm{p}g_i}{\Delta_a}$ with the transverse pump Rabi frequency $\Omega_p$. $\lambda_{12}=-\frac{g_1g_2N}{2\sqrt{2}\Delta_\textrm{A}}$ is the Raman coupling between one cavity and the other one and can be controlled via the atomic detuning $\Delta_\textrm{A}$. The atomic cloud acts as a dispersive medium in the cavity and changes the effective cavity lenght. The resulting dispersive shift of the cavity resonance is $N g_i^2/(2\Delta_a)\ll\Delta_i$. It is similar for both cavities and we absorbe it into $\Delta_i$. Higher order terms can be discarded for our experimental parameters. For our experimental parameters, collisional interactions between the atoms result only in a small overall energy shift. Since the vacuum Rabi couplings $g_i$ of the two cavities are slightly different, the BEC is aligned at the position where we achieve equal effective vacuum Rabi couplings $\tilde{g}_i$ of the two cavities due to the convolution with the mode profiles of the cavities. This position is slighlty off centered with respect to the mode crossing but ensures that we can use $g_0=\tilde{g}_1=\tilde{g}_2$ and a single coupling $\lambda=\lambda_1=\lambda_2$. We neglect the cavity decay rates which are not relevant in this part of the discussion. We also neglect effects due to the spatial phase of the transverse pump lattice relative to the cavity lattices. 

\noindent \textbf{Symmetries and phases} The Hamiltonian \eqref{eq:Hamiltonian} possesses two $\mathbb{Z}_2$ symmetries that can be broken independently. 
When photons are coherently scattered from the transverse pump into either of the cavities, a light field amplitude $\alpha_i$ in cavity $i$ becomes non-zero and can take either 0 or $\pi$ phase with respect to the transverse pump beam phase, as the electric field has to match the boundary conditions set by the cavity mirror. These two phases of the light field correspond in real space to atoms sitting on even or odd sites of the interference potential that is formed between the transverse pump beam and the cavity lattice.

\noindent \textbf{Low-energy mean field expansion}
Here we show an exact mapping of our microscopic model on a Landau free energy that captures the phenomenology of the phase diagram of two intertwined order parameters. Starting from the Hamiltonian $\hat{\mathcal{H}}$ we restrict our description to the low-energy momentum modes $\hbar\omega_-$ which are highlighted in Fig.~\ref{fig:momenta}:
\begin{equation}
\begin{aligned}
\hat{\mathcal{H}}_{\textrm{low-energy}} =&\sum_{i=1,2}\Bigl[-\hbar\Delta_i\hat{a}^{\dagger}_i\hat{a}_i+\hbar\omega_{-}\hat{c}^{\dagger}_{i-}\hat{c}_{i-}\\
&+\frac{\hbar\lambda}{\sqrt{N}}\Bigl(\hat{a}^{\dagger}_i+\hat{a}_i\Bigl)\Bigl(\hat{c}^{\dagger}_{i-}\hat{c}_{0}+h.c.\Bigl)\Bigl]\\
&+\frac{\hbar\lambda_{12}}{\sqrt{N}}\Bigl(\hat{a}^{\dagger}_1\hat{a}_2+\hat{a}^{\dagger}_2\hat{a}_1\Bigl)\Bigl(\hat{c}^{\dagger}_{12-}\hat{c}_{0}+h.c.\Bigl)\\
&+\hbar\omega_{-}\hat{c}^{\dagger}_{12-}\hat{c}_{12-}.
\end{aligned}
\label{eq:Hamiltonian low energy}
\end{equation}
We go to a mean-field description of this Hamiltonian by taking the average values $\langle\hat{a}_i\rangle=\alpha_i$ and $\langle\hat{c}_{i-}\rangle=\psi_i$ and $\langle\hat{c}_{12-}\rangle=\psi_{12}$. The mean-field Hamiltonian then becomes:
\begin{equation}
\begin{aligned}
\mathcal{H}_{\textrm{MF}} =\sum_{i=1,2}\Bigl[&-\hbar\Delta_i\alpha_i^2+\hbar\omega_{-}\psi_i^2+\frac{4\hbar\lambda_i}{\sqrt{N}}\alpha_i\psi_i\psi_0\Bigl]\\
&+\frac{4\hbar\lambda_{12}}{\sqrt{N}}\alpha_1\alpha_2\psi_{12}\psi_0+\hbar\omega_{-}\psi_{12}^2.
\end{aligned}
\label{eq:Hamiltonian mean field}
\end{equation}

\noindent \textbf{Effective potential for the intertwined phase}
To derive the Landau potential or free energy $\mathcal{F}$ that we show in the main text (see Fig.~\ref{intertwined:admixture}), we use the steady state solutions of the atomic fields $\psi_i$ and $\psi_{12}$ to reduce $\mathcal{H}_{\textrm{MF}}$ to a function of the photon field amplitudes $\alpha_1$ and $\alpha_2$. Therefore we set $\partial\mathcal{H}_{\textrm{MF}}/\partial\psi_i=0$ and obtain $\psi_i=-2\sqrt{N}\lambda_i\alpha_i\psi_0/\omega_-$ for $i$=1,2 and $\psi_{12}=-2\sqrt{N}\lambda_{12}\alpha_1\alpha_2\psi_0/\omega_-$. The atomic fields respect the normalization condition $\psi_0^2=N(1-\psi_1^2-\psi_2^2-\psi_{12}^2)$. Substituting these expressions into $\mathcal{H}_{\textrm{MF}} $ and keeping terms up to quartic order in $\alpha_1$ and $\alpha_2$, we obtain the expression for the free energy:
\begin{equation}
\begin{aligned}
\mathcal{F}(\alpha_1,\alpha_2;&r_1,r_2,g,\gamma_{12})/\hbar =\\&-\Delta_1\Bigl(1-\frac{\lambda^2}{\lambda_{1,\textrm{crit}}^2}\Bigl)\alpha_1^2-\Delta_2\Bigl(1-\frac{\lambda^2}{\lambda_{2,\textrm{crit}}^2}\Bigl)\alpha_2^2\\
&+\frac{16\lambda^4}{\omega^3}(\alpha_1^2+\alpha_2^2)^2-\frac{4\lambda_{12}^2}{\omega}\alpha_1^2\alpha_2^2
\end{aligned}
\label{eq:Landau}
\end{equation}
where we have set $\lambda_{i,\textrm{crit}}=\sqrt{-\Delta_i\omega/4}$. This expression of the free energy is identical to the one presented in the main text where $r_i=-\Delta_i\Bigl(1-\frac{\lambda^2}{\lambda_{i,\textrm{crit}}^2}\Bigl)$, $g=16\lambda^4/\omega^3$ and $\gamma_{12}=4\lambda_{12}^2/\omega$. Note that the detuning imbalance is $\bar{\Delta} \equiv \Delta_2-\Delta_1 = r_1-r_2$. The phenomenology of the phase diagram, which is experimentally measured, can be reproduced by this free energy $\mathcal{F}$. Terms higher than the quartic order should be included to calculate the position of the phase boundary of the intertwined phase. The ground state of the system for given parameters $r_i$, $g$ and $\gamma_{12}$ can be found by minimizing $\mathcal{F}$ with respect to both $\alpha_1$ and $\alpha_2$. In the case where $\gamma_{12}=2g$, the free energy $\mathcal{F}$ separates,
\begin{equation*}
\begin{aligned}
\mathcal{F}(\alpha_1,\alpha_2;&r_1,r_2,g,\gamma_{12})/\hbar =r_1\alpha_1^2 + g\alpha_1^4+r_2\alpha_2^2 + g\alpha_2^4\\
&=\mathcal{F}(\alpha_1;r_1,g)/\hbar +\mathcal{F}(\alpha_2;r_2,g)/\hbar.
\end{aligned}
\end{equation*}
In this case the phase boundaries of the two ordered phases SO1 and SO2 do not influence each other through the phase diagram as it is measured in Fig.~2{\bf c} of the main text.

\bibliography{bib/references}

\appendix

\noindent \textbf{Acknowledgements} We thank Eugene Demler, Sarang Gopalakrishnan, Awadhesh Narayan, Yulia E. Shchadilova and Nicola Spaldin for insightful discussions. We thank Davide Dreon for careful reading of the manuscript and Xiangliang Li for experimental assistance. We acknowledge funding for the SBFI Horizon2020 project QUIC (grant agreement 641122), the Horizon2020 European Training Network ColOpt (grant agreement 721465), and SNF support for NCCR QSIT and DACH project `Quantum Crystals of Matter and Light' .
 
\noindent \textbf{Author Contributions} All authors contributed extensively to the work presented here.
 
\noindent \textbf{Author Information} The authors declare no competing financial interests. Correspondence and request for materials should be addressed to T.E. (esslinger@phys.ethz.ch).
 
\end{document}